\documentclass[a4paper,10pt]{article}
\usepackage{url}
\usepackage{helvet}
\usepackage{courier}
\usepackage{graphicx}
\usepackage{algorithm}
\usepackage{algpseudocode}

\title{Lateral Movement Detection Using User Behavioral Analysis}
\begin{document}
\author{Deepak Kushwaha$^{a}$,
        Dhruv Nandakumar$^{a}$,\\
        Akshay Kakkar$^{a}$,
        Sanvi Gupta$^{a}$,\\
        Kevin Choi$^{a}$,
        Christopher Redino$^{a}$,
        Abdul Rahman$^{a}$,\\
        Sabthagiri Saravanan Chandramohan$^{a}$,
        Edward Bowen$^{a}$,\\
        Matthew Weeks$^{a}$,
        Aaron Shaha$^{a}$,
        Joe Nehila$^{a}$\\
        \small $^{a}$Deloitte \& Touche LLP \\
        \small $^{*}$Corresponding author: kevchoi@deloitte.com \\
}
\maketitle

\begin{abstract}
Lateral Movement refers to methods by which threat actors gain initial access to a network and then progressively move through said network collecting key data about assets until they reach the ultimate target of their attack. Lateral Movement intrusions have become more intricate with the increasing complexity and interconnected nature of enterprise networks, and require equally sophisticated detection mechanisms to proactively detect such threats in near real-time at enterprise scale. In this paper, the authors propose a novel, lightweight method for Lateral Movement detection using user behavioral analysis and machine learning. Specifically, this paper introduces a novel methodology for cyber domain-specific feature engineering that identifies Lateral Movement behavior on a per-user basis. Furthermore, the engineered features have also been used to develop two supervised machine learning models for Lateral Movement identification that have demonstrably outperformed models previously seen in literature while maintaining robust performance on datasets with high class imbalance. The models and methodology introduced in this paper have also been designed in collaboration with security operators to be relevant and interpretable in order to maximize impact and minimize time to value as a cyber threat detection toolkit. The underlying goal of the paper is to provide a computationally efficient, domain-specific approach to near real-time Lateral Movement detection that is interpretable and robust to enterprise-scale data volumes and class imbalance.
\end{abstract}

\section{Introduction}
The increasing size, complexity, and interconnected nature of enterprise networks is accompanied by an increasingly complex attack surface. Malicious actors leverage this connected nature by infiltrating weaker parts of networks and then using elevated credentials to move between interconnected hosts in the search for sensitive data or other high value assets. Lateral Movement (LM) attacks are a key differentiator responsible for many of today’s Advanced Persistent Threats (APTs) \cite{crowdstrike}. Detecting instances of malicious LM is paramount to maintaining effective enterprise security posture and being able to proactively mitigate breaches that could cost hundreds of millions of US dollars \cite{deloitte-cyberattack}. 

In this paper, we introduce a method to detect LM in remote authentication events by utilizing user-focused feature engineering created specifically to capture LM behaviors and supervised classifiers trained to accommodate for highly imbalanced datasets. The goal of the methodology is to enable robust and interpretable LM detection in near real-time using lightweight classifiers and a highly generalized feature engineering pipeline while simultaneously prioritizing extremely low False Positive Rates (FPR) to prevent alert fatigue of end users. The key contributions of this paper are:
\begin{itemize}
\item Providing a novel feature set designed to specifically identify LM behaviors in user authentication data using user-based behavioral analysis.
\item Benchmarking supervised classifier performance on our novel feature set on a realistic,  and highly imbalanced dataset.
\item Exploring a framework for model interpretability that allows security operators to effectively understand model decision criteria on a per-example basis.
\end{itemize}

The paper begins with a literature review to explore key efforts toward LM detection that exist today, followed by sections that discuss our feature engineering, modeling approach, evaluation, interpretability, and finally a conclusion and discussion of future work.

\section{Background}

The objective of this section is to outline previous work conducted in the identification of LM in enterprise networks using Machine Learning (ML) based approaches. Most of the approaches seen in literature thus far belong predominantly to either rules-based methods or unsupervised learning methods with a few studies utilizing supervised ML. One such rule-based study was conducted by \cite{bowman2020detecting} wherein the authors classified all first time login events between a user and device as anomalous. Similarly, other rules-based approaches could also be conducted such as classifying all failed login events as anomalous. However, the impact of scale and FPR are readily apparent when using the above methods, especially as the size of enterprise networks grows. Utilizing more advanced ML based methods would not only drastically improve efficiency and reduce FPR but also allow for more complex APTs to be captured from authentication logs. 

Unsupervised learning approaches to LM detection allow for detection of more complex anomalous behavior in authentication data as compared to rule-based approaches. The notion is to be able to group/cluster users based on patterns of similar behavior to identify if any particular users are anomalous insofar as they do not belong to a large enough cluster. \cite{bohara2017unsupervised} implemented a graph-based approach to LM detection utilizing Network and Host Process logs, wherein the authors extracted graph-based features from the aforementioned logs and implemented Principal Component Analysis (PCA) \cite{wold1987principal} and K-means clustering \cite{hartigan1979algorithm} to identify outliers in the dataset. \cite{bowman2020detecting} also utilized an unsupervised graph-based technique using authentication data by implementing a logistic regression based method to predict low-probability links between assets. While this approach demonstrated strong results with an FPR of only 0.9\%, the actual dataset used for evaluation was significantly smaller than those expected in enterprise networks. \cite{semi-sup} also utilized host-communication graphs as a basis for feature extraction using a denoising autoencoder followed by a K-Nearest-Neighbors (KNN) classifier \cite{1056403} for LM classification. The approach produced very strong results with a precision of 91\% and an FPR of 0.01\%, but training and evaluation was conducted on a dataset wherein 6.8\% of the data was malicious and performance is anticipated to degrade as the class imbalance intensifies to enterprise scale. Furthermore, graph based approaches in general may be harder for end-users (security operators) to interpret, making results significantly less usable \cite{du2018towards}. It is also worth noting that the efficacy of unsupervised methods depends heavily on the characteristics of the underlying data and require continuous per-network hyperparameter tuning to ensure that the results of the model are meaningful - which make such models less feasible to deploy in enterprise networks efficiently.

Supervised ML approaches tend to overcome the hurdle of deployment efficiency by utilizing a pre-trained model for LM detection on Information Technology (IT) networks. Supervised methods can also be trained to account for heavy class imbalance and complex data distributions. \cite{bai2021rdp} utilized a variety of supervised ML methods on user authentication and host process data from the Los Alamos National Laboratory (LANL) unified and comprehensive datasets \cite{kent2015comprehensive, lanl-data} including Logistic Regression \cite{hosmer2013applied}, Random Forest (RF) classifiers \cite{breiman2001random}, LogitBoost (LB) \cite{10.1214/aos/1016218223}, and Light Gradient Boosting Machine (LightGBM) \cite{ke2017lightgbm} methods. The results of the study demonstrated very strong performance with over 99\% accuracy, precision, and recall but the models were trained and evaluated on a dataset containing 56,837 events, which will likely not be representative of enterprise scale class imbalances. \cite{uppstromer2019detecting} highlights the performance of different supervised classifiers such as Decision Tree (DT) \cite{song2015decision}, RF classifier, KNN, Support Vector Machines (SVM) \cite{10.1145/130385.130401} on a more imbalanced dataset containing 1\% malicious activity with the RF model achieving 98\% precision and 86\% recall, which is already significantly lower than in \cite{bai2021rdp}.

Clearly, the development of a model that can operate with extremely high precision, recall, and low FPR using highly imbalanced data representative of enterprise networks is imperative. Furthermore, it is also important for the said model to be interpretable to maximize its usability while simultaneously being computationally efficient. Through the rest of this paper, we will explore the creation of such a model.

\section{Methodology}

The primary motivation for this work is to be able to build a supervised ML model that can identify LM behaviors in enterprise networks effectively. This entails being able to capture malicious authentication activity in near real-time reliably while simultaneously avoiding alert fatigue on security operators by reducing the number of false positives. We believe that we can achieve strong results with user-based behavioral analysis of authentication and process events using domain-specific features designed to mimic indicators that security operators use while investigating potential LM events. Herein, we will define and describe a comprehensive methodology for creating such domain-specific features while minimizing computational overhead. The feature engineering aims to construct data such that all normal user/authentication data will have similar feature values which will also vary significantly from feature values exhibit by malicious actors. In subsequent sections we are going to explore the datasets we utilized for modeling, our feature engineering process, and the complete viability of derived features using Exploratory Data Analysis (EDA).

\subsection{Dataset}
In this paper, we used the LANL Comprehensive, Multi-Source Cyber-Security Events dataset \cite{kent2015comprehensive} to train and evaluate our models and approach. It is a high-quality, publicly available assessment dataset with labelled cases targeting LM behavior. It contains real network traffic data from a live exercise at LANL, where red team hackers ran an attack campaign on the network to simulate an adversarial event. The dataset contains security logs for 58 days, with only 0.0000713\% of all occurrences flagged as malicious, resulting in an highly imbalanced dataset. However, we believe that this is representative of realistic probabilities of finding such events in enterprise data and consequently prioritized modeling and feature engineering efforts to maintain performance despite the imbalance.

The LANL dataset includes network flow telemetry, authentication logs, system process execution logs, and Domain Name System (DNS) logs. However, we only utilize the authentication and process logs from the dataset which contains data\footnote{In some cases, missing data fields are indicated using '?'} as described below. 

The process dataset contains 62,947 events with each event containing the following fields:
\begin{itemize}
    \item \textbf{time}: The time at which the process event occurred, measured in seconds since the first event in the dataset. 
    \item \textbf{user@domain}: The user identity and domain of the user who’s computer executed said process.
    \item \textbf{computer}: The identifier for the device that executed said process.
    \item \textbf{process name}: The name of the process that was executed.
    \item \textbf{start/end}: Flag indicating whether a process execution began or terminated.
\end{itemize}

The authentication dataset contains 1.05 billion authentication events, which are comprised of 9 fields:
\begin{itemize}
\item \textbf{time}: The time at which the authentication event occurred, measured in seconds since the first event in the dataset. 
\item \textbf{source user@domain}: The source user logging in and the associated domain the user is using.
\item \textbf{destination user@domain}: The username and the associated domain after the user has performed an authentication event.
\item \textbf{source computer}: The origin computer used for the authentication event.
\item \textbf{destination computer}: The destination computer where the authentication event terminates.
\item \textbf{authentication type}: The type of Windows-based authentication used. It contains values like \textit{NTLM}, \textit{Kerberos}, etc.
\item \textbf{logon type}:  The type of authentication occurring includ- ing cross-network authentication, an interactive keyboard session, a batch event, a system service, a screen saver lock or unlock, and several others.
\item \textbf{authentication orientation}: This includes values indicating whether the event is for granting a \textit{Kerberos}, \textit{TGT} or \textit{TGS}, a log on or log off event, or an authentication credential mapping. 
\item \textbf{success/failure}: A flag indicating whether the authentication event is successfully completed or failed. Failure could happen for several reasons including incorrect passwords, a locked out account, or an authorization failure.
\end{itemize}

\subsection{Feature Engineering}

While the LANL dataset is, indeed, realistic in scale and quality, we believe that domain-specific feature engineering is necessary to facilitate ML models to distinguish between malicious and benign events. These features are efficient in simultaneously capturing user usage patterns and also the complicated anomalies that emerge within the network due to an adversary’s intrusion, further resulting in an LM attack. The following features are derived from the existing fields in the authentication and process event logs and are designed specifically to identify LM behaviors in a manner similar to how Security Operations Center (SOC) personnel would identify LM. Furthermore, given LM occurs on a per-user basis, our feature engineering is focused primarily on features that aim to capture information about user-behavior while avoiding the effect of authentication activity carried out by other users within the network in the process. This has the added benefit of reducing the computational overhead of feature engineering in enterprise networks while still maintaining consistently strong performance as an anomaly detector. The following features were designed based on feedback and rationale provided by our organization's cyber Subject Matter Experts (SME) team:

\begin{itemize}
\item \textbf{Remote}: Whether the event is a remote login or a local login. A remote login event will have an authentication activity originating from a source computer and terminate on a different destination computer. While interactive events will be used during feature engineering, only remote events will be considered during model training. This is due to the fact that LM events will always have to be remote in nature. Successive interactive (physical) logins are not, by definition, LM and can be classified as benign \textit{a priori}.
\item \textbf{Duration}: Duration of login session between a successive login and logout by a user. For every session, after the logoff event, the duration is calculated by identifying the most recent login event from the user to the same source computer. Intrusion activity may result in very short, near instantaneous sessions when used as part of an automated attack, while interactive operators using a login session to maintain remote access to a system may extend the session much longer than ordinary activity.
\item \textbf{Was source logged on}: Whether source user account is already logged into source computer at the time of performing a remote authentication event. This can be identified as any interactive login event within the past 24 hours with no logout event afterward. This specific feature can capture behavior where a malicious actor is trying to perform a remote login on a destination within the network, without having any record of interactive source login on the same computer, a scenario that often happens when employing stolen credentials.
\item \textbf{How long ago}: In incidents where the source computer is logged on interactively within past 24 hours, this feature captures the time span between the current remote event and the last interactive login from same source computer. It is calculated by finding the time difference between the most recent interactive login event and the current remote event from the same source computer.
\item \textbf{How many other interactive logins}: The number of other active interactive logins to other computers that this user had at the time of the remote event. Typically, a source user is expected to be logged in from their credentials to 1 or at most 2 source computers interactively and having a higher count higher could indicate suspicious behavior.
\item \textbf{Was process run}: A flag indicating whether a new process gets executed on the destination computer within 60 seconds of a remote login. This characteristic identifies behaviors where an adversary can start a process soon after performing LM and gaining a foothold of destination computer.
\item \textbf{Previous login fraction}: Fraction of source account’s previous logins within an hour of the current time of day (i.e., business hours). The feature captures behavior of attackers where the authentication event occurs out of the typical business hours for the source user. This feature will have abnormal values if, for example, the credential of a user is used to login a computer around midnight, who usually logs on in the first half of the day.
\item \textbf{Number of failed}: Number of failed remote logins from the same source computer within past 1 hour. In many instances, malicious LM activity occurs with reused credentials the attackers obtained elsewhere or guessed and there is a high probability of failed events as the attacker tries other passwords before finding the correct one. A remote event captured after multiple failed attempts by any user on same source computer could be more indicative of a malicious actor.
\item \textbf{Fraction from same box}: This captures a user’s inherent behavior of logging in from the same source computer. It is measured as the fraction of source account’s previous logins have been from the same source computer. Any deviation of user’s behavior of logging in from other source computer will be captured and represented by low percentage.
\end{itemize}

\subsection{Exploratory Data Analysis}
After completing the process of feature engineering, our goal is to perform specific EDA and determine how valuable domain-specific features are in defining a clear separation boundary between malicious and benign remote login activity. To depict the variance in feature values for malicious and non-malicious users, standard statistical measures with aggregated values for each feature are compared. The goal of this statistical analysis is to decide whether the new features provide visibility in users having malicious authentication events and differentiate them from benign users only based on statistical values. As a result, the investigation revealed that aggregated values for individual user data had less influence than individual authentication events and hence, future techniques will only employ individual authentication occurrences to construct an ML solution. That is, the ML models in this paper are trained on each of the authentication events for each user individually rather than in an aggregated form.

\begin{figure}[htp]
    \centering
    \includegraphics[width=8cm]{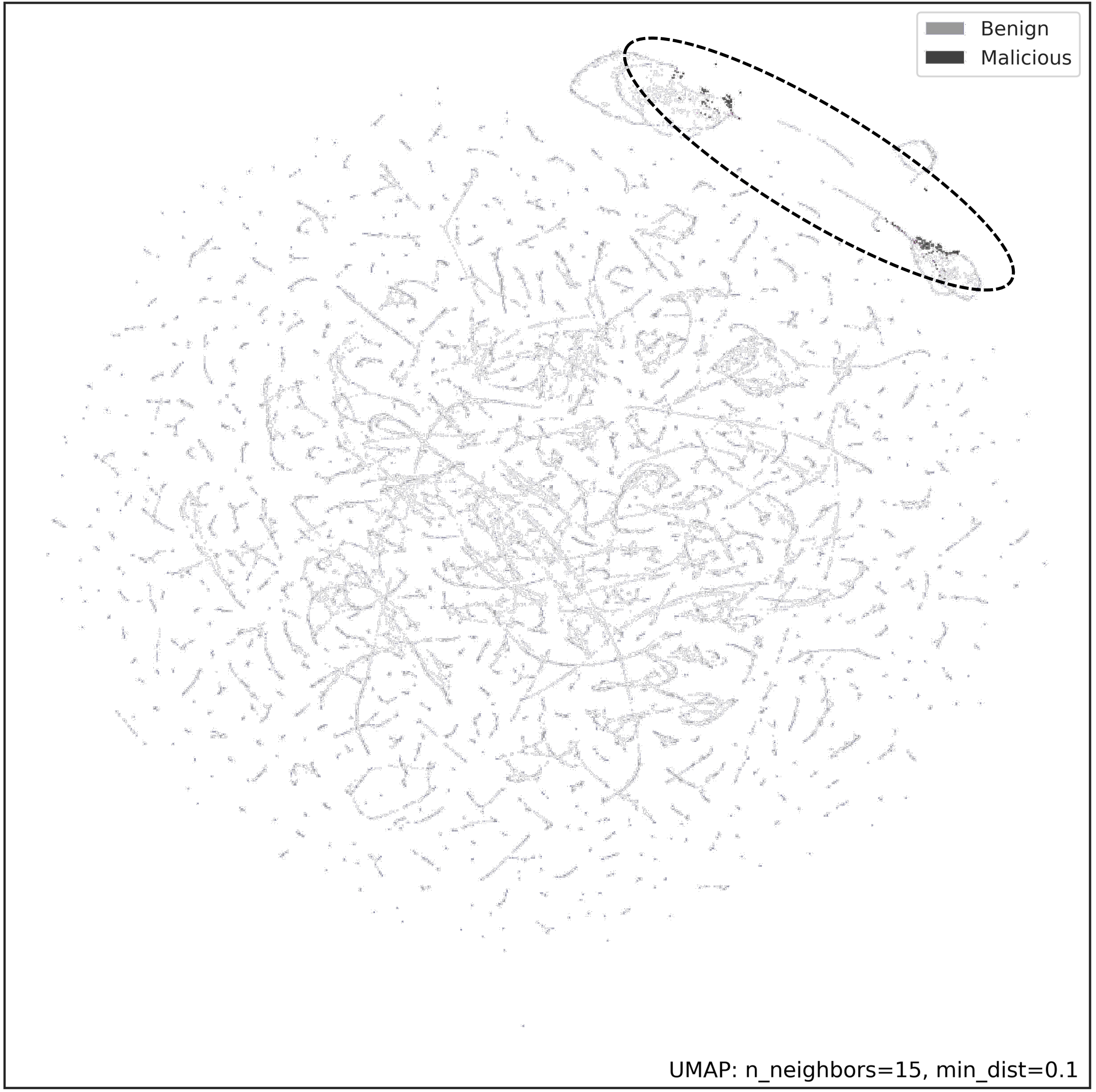}
    \caption{UMAP with a distribution of benign and malicious events}
    \label{fig:UMAP_}
\end{figure}

Next, we explore spatial separation between benign and malicious authentication events in the LANL dataset with our new features to understand if they can be clearly discerned by an ML model. Given the scale of data, it is computationally infeasible and impractical to visualize all data points. Hence, the visualization is performed by using the Uniform Manifold Approximation and Projection (UMAP) technique \cite{umap} for individual authentication events from a random stratified subset of the data having 100,000 authentication data points (574 malicious events + 99,426 normal events) in Fig. \ref{fig:UMAP_}. The normal authentication events are shown in lighter shade of grey, while the malicious ones representing LM are shown in black.

As shown Fig. \ref{fig:UMAP_}, it is evident that benign and malicious data are fairly discernible in a latent space using the features we constructed, which is a promising result. From here, we can focus on using this difference in data distributions between classes to create a classifier that distinguishes malicious from non-malicious activity. A significant challenge in designing a binary class classifier is handling the extreme data imbalance and approaches to do so will be explored henceforth.

\section{Experimental Design}

Our feature set is bench-marked on supervised ML models that are designed as binary classifiers. This is done in an effort to facilitate near real-time detection using authentication and process data with minimal hyper-parameter optimization post training. The goal of these models will be to look at each event independently and classify them as malicious or non-malicious based on the values of user-behavior-based domain-specific features. Furthermore, as discussed in Methodology section, interactive (non-remote) authentication events are not used during model training or evaluation and are discarded after being used for feature generation. This section will next discuss our data preparation process, how we train using imbalanced data and the bench-marking models we apply.

\subsection{Data Preparation}
Post feature engineering Eight features are selected for model training, seven of which are derived and one from raw data fields: \textit{remote, was source logged on, how long ago, how many other interactive logins, was process run, previous login fraction, number of failed, fraction from the same box} and \textit{authentication type} respectively. Two of the nine features derived through feature engineering were dropped: \textit{remote} and \textit{duration}. Since only remote events are used to train the model, this binary characteristic is unnecessary for training. Similarly, the duration value is only generated upon session logoff, whereas the goal of this method is to identify anomalies in real time for every remote login event. The authentication type, which is a categorical raw feature, needed to be transformed using label encoding. All other features are included without any modifications. Training and validation sets are captured from the existing data in a ratio of 80:20. The data split has been very carefully achieved on a user basis and stratified by the number of malicious activities within those user authentication activities. Therefore, all of the authentication events of an individual user reside in either the training set or the test set. This distinction is important to accurately validate the model to new users and determine that our models can generalize effectively. It is also important to note here that aside from focusing on only remote events, the training and testing data are not over-sampled or under-sampled in any way and the original class imbalance is maintained to make sure our model performance accurately reflects real-world performance. After excluding local interactive and logoff events, the training dataset contains a total of 263,716,042 authentication events, with only 485 events tagged as malicious. The test set contains 66,530,377 events, out of which 89 events are labelled as malicious.

\subsection{Model Selection and Imbalanced Classification}
Given that the objective of our models will be to accurately classify LM attempts in enterprise network data, we need to choose modeling approaches that can accommodate both complex non-linear separations in data as well as extreme imbalance. Consequently, we chose to implement two in- dependent models to classify LM events: Extreme Gradient Boosting (XGBoost) \cite{chen2016xgboost} and Fully Connected Deep Neural Networks (FCDN) \cite{hecht1992theory}. We selected XGBoost as our first model due to the fact that DTs generally perform well with imbalanced data and ensemble techniques with DTs (RF, XGBoost, etc.) almost always outperform singular DTs \cite{zhang2012ensemble}. FCDNs were utilized because of their ability to perform well in highly complex decision spaces. However, we augmented the training processes for both models with cost-sensitive training that utilizes class weighting while computing model loss to assign examples of the minority class a higher weight. Furthermore, we also utilized fit for purpose evaluation metrics that gave us a clear understanding of the performance of our models on classifying imbalanced data: Recall scores and FPR.

\subsection{Model Architecture}

\subsubsection{The XGBoost Approach}
XGBoost is a scalable tree ensemble boosting system algorithm, which is used widely by data scientists to achieve state-of-the-art results on many ML tasks \cite{chen2016xgboost}. It facilitates training capabilities for in-memory as well as out of memory datasets. The tree-based algorithm can ingest data for training without the need of feature scaling/normalization and are treated as one of the best suited algorithms for imbalance classification problem in statistical ML domain. Given the size of the training data, a distributed version of XGBoost is used to train and validate the models. To handle the highly imbalanced distribution of malicious and benign activities within the training data, the parameter \textit{scale\_pos\_weight} which controls the balance of positive and negative weights is modified. As the class imbalance is quite high, we used a more conservative value (\textit{scale\_pos\_weight}=1000) for this parameter to strike a balance between recall scores and FPR. This scaling parameter was identified by rapid hyper-parameter tuning and simultaneously maintaining a balance between recall scores and FPR.

\subsubsection{The Neural Network Approach}
An FCDN is trained as a binary classifier using a feed-forward neural network followed by a single node output, which outputs the probability of an authentication event being an instance of LM. Hyperparameter tuning and optimization is conducted to optimize various parameters in the model including activation functions, dropout probabilities, layer depth and width. Finally, the utilized model architecture comprises 4 layers, with an input layer of  \textit{N} neurons that handles the input data format following data transformation and feature selection. The two subsequent hidden layers have 14 and 7 neurons respectively with Rectified Linear Unit (ReLU) activation functions, while the output layer has one neuron which is sigmoid activated. The model is trained using a custom loss function namely, Weighted Binary Cross Entropy (WBCE) loss. WBCE loss is used to apply the notion of class weighting and penalizing in order to address class imbalance. This loss function penalizes the class with more data points by using the weights associated with each class. Below is a custom weighted binary cross entropy loss function, with a value of 1000 applied to the class weights of the minority class in order to balance the classifier.

\begin{algorithm}
\caption{Weighted Binary Cross Entropy Loss}\label{alg:WBCE_Loss}
\begin{algorithmic}
\Require $weights \gets dictionary$
\Require $labels \gets y\_true$
\Require $logits \gets y\_pred$
    \State $loss \gets labels * -log(sigmoid(logits)) * weights + (1 - labels) - log(1 - sigmoid(logits))$
\Ensure $loss$
\end{algorithmic}
\end{algorithm}

\subsection{Experimental Setup}
The experiment results presented in this paper are the result of exhaustive feature selection. In order to attain the final results, the experimental technique employs a greedy approach to feature selection, selecting the most representative or significant feature from the remaining features at each iteration while evaluating the results. An identical set of training and testing data is utilized by both models however, the set of input features for each model differs depending on the feature selection process. The reported findings are best obtained after extensive hyper-parameter optimization and comprehensive feature selection. Given that feature generation for our training and evaluation datasets needs to be performed only once, features were pre-computed and stored on disc in a partitioned format for faster read access during training. For the XGBoost approach, model training and evaluation are performed in a distributed environment utilizing Dask \cite{rocklin2015dask} on a single machine. Given the quantity of the data, Dask-distributed XGBoost trains in parallel on all machine cores to provide the best performance results. FCDNs can handle huge amounts of data due to their support for in-memory, batch-based training and their training on Graphics Processing Unit (GPU) support for faster training; however, their overall training time is significantly longer than that of XGBoost. Both models were trained on instances having at least 256 GB of RAM and 24 GB of V-RAM.

\subsection{Success Criteria}
Given the highly imbalanced nature of the dataset and the highly severe nature of an LM attack, we optimize for two metrics in particular: higher recall to prioritize the detection of malicious activity and a lower FPR to reduce operator alert fatigue. Our success criteria were to achieve a recall greater than 70\% and a FPR less than 0.005\%. It should also be noted that predictions are treated as LM events if the probability of their prediction is greater than 50\%. 

\section{RESULTS AND DISCUSSION}

\begin{figure}[htp]
    \centering
    \includegraphics[width=8cm]{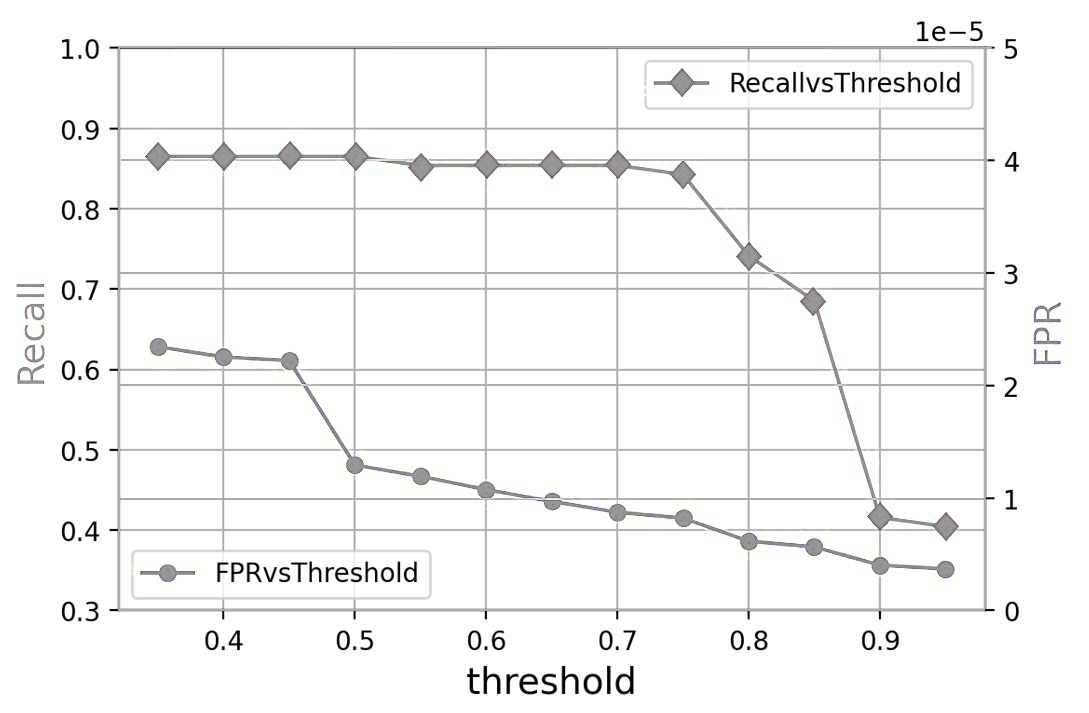}
    \caption{XGBoost Approach}
    \label{fig:XGB_curve}
\end{figure}

\begin{figure}[htp]
    \centering
    \includegraphics[width=8cm]{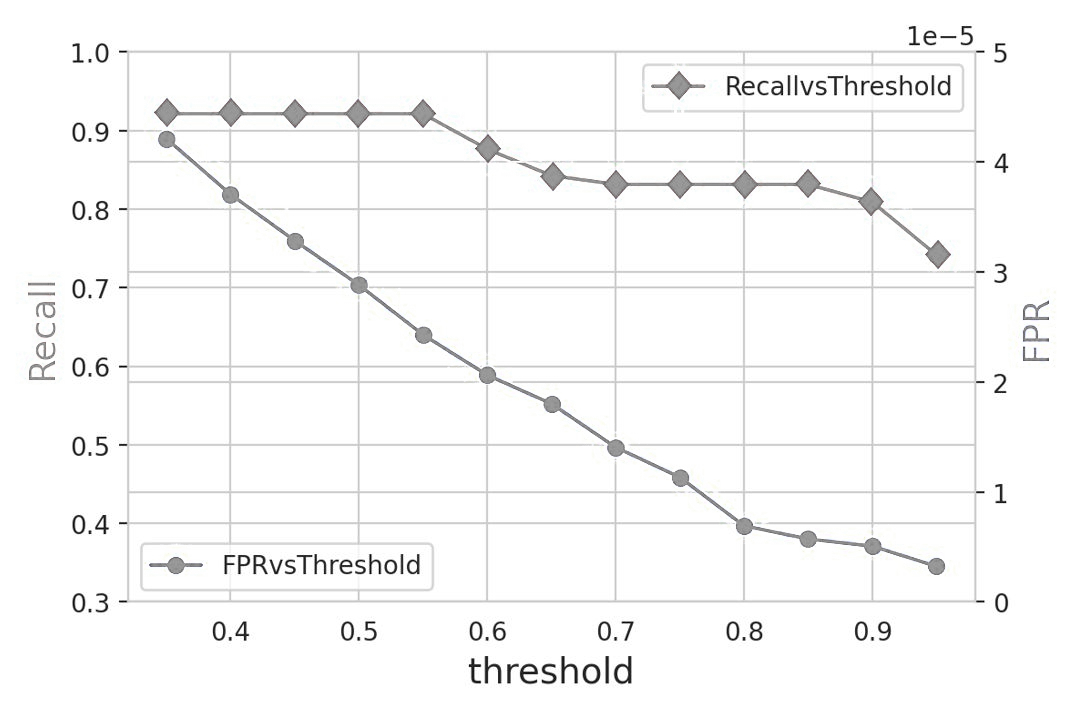}
    \caption{Deep Learning Approach}
    \label{fig:DL_curve}
\end{figure}

\subsection{Results}
The experiments with XGBoost achieved an average recall score or malicious activity detection score of 86.51\% and FPR of 0.0013\% for our experimental setups. The results obtained appear to maintain strong performance despite higher class imbalance compared to other work in  \cite{du2018towards, uppstromer2019detecting, bai2021rdp}, which makes the model more suitable for enterprise threat detection. The results achieved by the FCDN have a recall score of 92.13\% and an FPR of 0.0029\%. Both the models show a strong performance in predicting either of the classes with Recall vs Threshold and FPR vs Threshold as shown in  Fig. \ref{fig:XGB_curve} and Fig. \ref{fig:DL_curve} for XGBoost model and the FCDN respectively.

\begin{table}[h]
\begin{center}
\begin{tabular}{||c c c||}
 \hline
 Algorithm & Recall & FPR \\ [0.5ex]
 \hline\hline
 XGBoost & 86.51\% & 0.0013\%  \\[0.5ex]
 \hline
 FCDN & 92.13\% & 0.0029\%  \\[0.5ex] 
 \hline
\end{tabular}\caption{Recall and FPR scores achieved}
\label{table:1}
\end{center}
\end{table}

It is evident that the XGBoost results are comparatively better if lower FPRs are the priority. However, both models exhibit very low FPRs in general, with FCDNs providing good detective capabilities for LM in general. Consequently, the two models demonstrated here can be used in conjunction to provide stronger detection capabilities. Particularly, the XGBoost model can be utilized by SOC personnel where operators triage alerts due to their lower FPR and the FCDN can be utilized by threat hunters, where stronger recall would be the priority. These results also give us a promising indication that our domain-specific features provide a strong foundation for LM detection.

\section{Model Interpretability}
We believe that our models’ performance is only strengthened by the fact that the underlying feature set used for prediction is inherently explainable to security operators. Consequently, we set out to build a framework for model interpretability that can validate our assumptions about feature engineering ex post facto as well as provide per-sample explanations of model predictions which end users can utilize as trail-heads for threat hunts. In general, model interpretability aims at giving precise explanations on the outcome of the model and specific features which influence the model output either positively or negatively. We utilize the popular model explainability tool Shapley (SHAP) \cite{shap} for the detailed analysis of model outputs, which assigns SHAP values to features in order to accurately explain feature importance. It's novel components include:
\begin{itemize}
    \item The identification of a new class of additive feature importance measures.
    \item Theoretical results showing there is a solution in this class with a set of desirable properties.
\end{itemize}

SHAP values provide per-feature global and local interpretability and serve as the basis for our LM model interpretability.

\begin{figure*}
    \includegraphics[width=\textwidth]{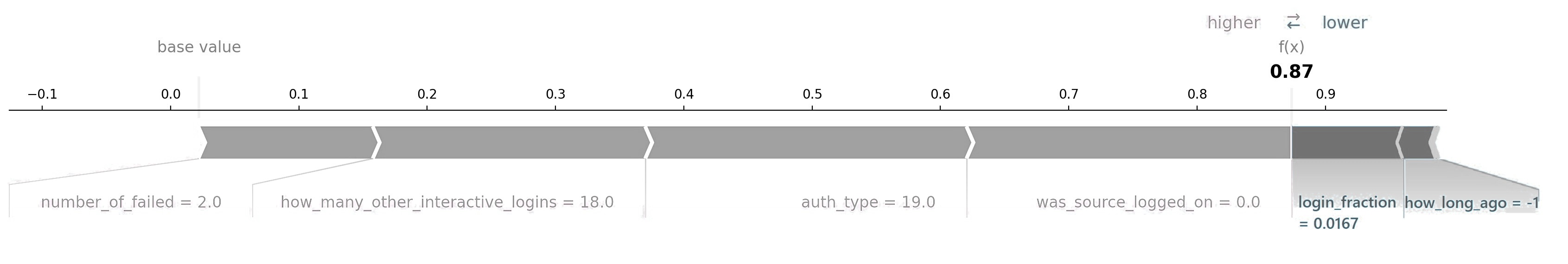}
    \centering
    \caption{Model Local Interpertability}
    \label{fig:local_interpretibility}
\end{figure*}

\subsection{Local Interpretability}
SHAP provides a defined value (Shapley value) for every feature employed while creating the model, which explains the positive or negative effect on the model output of malicious or non-malicious events and is only provided when a model prediction is malicious. This design choice was taken since benign activity would never be reported to security operators. The local interpretability as seen in our models allows threat hunters to understand why model determinations were made on a \textbf{per sample basis} as shown in Fig. \ref{fig:local_interpretibility}, in context to domain-specific features they can understand. Fig. \ref{fig:local_interpretibility} can be understood as follows :
\begin{itemize}
    \item \textbf{f(x)}: The output value is the prediction for that observation (Confidence score/Probability).
    \item \textbf{The base value}: The value that would be predicted if no features for the current output were known. i.e., it is the mean prediction confidence on the test dataset.
    \item \textbf{Directed arrow bars}: Features that push the prediction probability higher (to the right) are represented by arrow bars pointing to the right, and those pushing the prediction lower are shown with arrow bars pointing left.
\end{itemize}
Our hope is that these per-sample explanations of feature importance can help end users contextualize their threat investigations and ultimately lead to dramatically reduced time to detection as well as response.

\subsection{Global Interpretability}
\begin{figure}[htp]
    \centering
    \includegraphics[width=8cm]{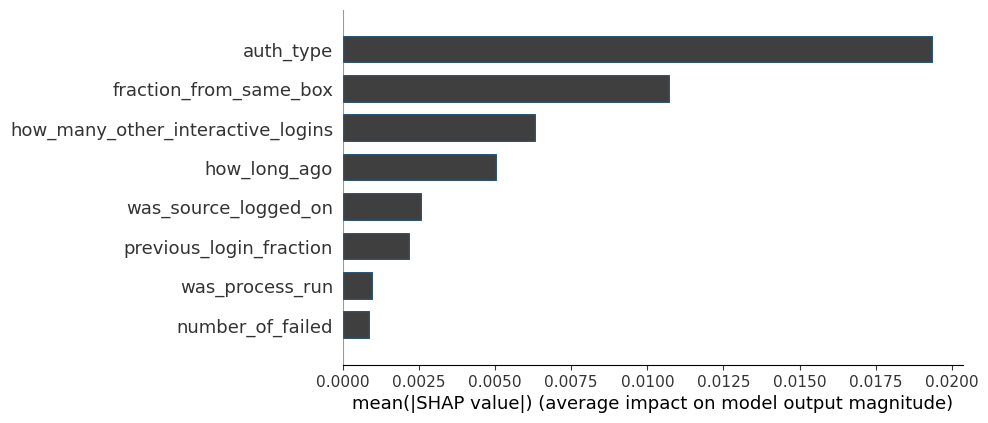}
    \caption{Model Global Interpretability}
    \label{fig:global_interpretation}
\end{figure}

Global interpretability describes the independent feature importance that influence our models across all training samples. This enables us to evaluate how effective our domain-specific features are at identifying LM .Furthermore, explore feature importance to determine if they correlate with the significance a security operator would assign them during threat hunts. Fig. \ref{fig:global_interpretation} provides an overview of the aggregate global feature importance as computed by our models.

As shown in Fig. \ref{fig:global_interpretation}, all of the domain-specific features proved to be important in providing discriminatory power to our models during prediction. Furthermore, review from our organization's cyber SME team confirmed that the feature importance learned by our models are in line with importance they would have assigned to the features during a threat hunt, which is a promising indicator of model performance and interpretability.

\section{Conclusion and Future Direction}

We propose implementing a novel method for LM detection using user-based behavioral analysis and domain-specific feature engineering using authentication and process logs. Our feature engineering approach was also bench-marked using two supervised ML approaches to confirm their efficacy, and it is evident that the features described provide stronger performance than other approaches seen in literature despite maintaining realistic, high class-imbalance while supporting near real-time, versatile detection capabilities. We provide a robust, computationally efficient methodology for LM detection with high recall and very low FPR.  We have further demonstrated that utilizing both deep learning and tree-based approaches in a security workflow provides the added benefit of allowing threat hunters and SOC analysts to utilize different models for different priorities. This approach also provides the added benefit of being explainable to end users because of the security-first of the underlying features themselves. We utilize this aspect of our data and models to build a robust interpretability framework for LM detection that confirms our feature engineering assumptions and provides a per-sample explanation of results to end users for threat hunting. 

Although our LM detectors maintain strong performance on imbalanced LANL dataset, the dataset only captures malicious activity on a Windows based system. Consequently, we would like to explore modeling approaches from various other operating systems such as Linux. We also plan to benchmark our models on multiple enterprise networks to measure and discern the models’ ability to generalize to novel networks without retraining. Furthermore, we would like to build and incorporate real-time end user feedback into our model hyperparameter optimization to allow them to adapt to a continually changing enterprise environment.

\bibliographystyle{apalike}
\bibliography{main.bib}

\begin{thebibliography}{}

\bibitem[Bai et~al., 2021]{bai2021rdp}
Bai, T., Bian, H., Salahuddin, M.~A., {Abou Daya}, A., Limam, N., and Boutaba,
  R. (2021).
\newblock Rdp-based lateral movement detection using machine learning.
\newblock {\em Computer Communications}, 165:9--19.

\bibitem[Bohara et~al., 2017]{bohara2017unsupervised}
Bohara, A., Noureddine, M.~A., Fawaz, A., and Sanders, W.~H. (2017).
\newblock An unsupervised multi-detector approach for identifying malicious
  lateral movement.
\newblock In {\em 2017 IEEE 36th Symposium on Reliable Distributed Systems
  (SRDS)}, pages 224--233.

\bibitem[Boser et~al., 1992]{10.1145/130385.130401}
Boser, B.~E., Guyon, I.~M., and Vapnik, V.~N. (1992).
\newblock A training algorithm for optimal margin classifiers.
\newblock In {\em Proceedings of the Fifth Annual Workshop on Computational
  Learning Theory}, COLT '92, page 144–152, New York, NY, USA. Association
  for Computing Machinery.

\bibitem[Bowman et~al., 2020]{bowman2020detecting}
Bowman, B., Laprade, C., Ji, Y., and Huang, H.~H. (2020).
\newblock Detecting lateral movement in enterprise computer networks with
  unsupervised graph {AI}.
\newblock In {\em 23rd International Symposium on Research in Attacks,
  Intrusions and Defenses (RAID 2020)}, pages 257--268, San Sebastian. USENIX
  Association.

\bibitem[Breiman, 2001]{breiman2001random}
Breiman, L. (2001).
\newblock Random forests.
\newblock {\em Machine learning}, 45(1):5--32.

\bibitem[Chen et~al., 2018]{semi-sup}
Chen, M., Yao, Y., Liu, J., Jiang, B., Su, L., and Lu, Z. (2018).
\newblock A novel approach for identifying lateral movement attacks based on
  network embedding.
\newblock In {\em 2018 IEEE International Conference on Parallel Distributed
  Processing with Applications, Ubiquitous Computing Communications, Big Data
  Cloud Computing, Social Computing Networking, Sustainable Computing
  Communications (ISPA/IUCC/BDCloud/SocialCom/SustainCom)}, pages 708--715.

\bibitem[Chen and Guestrin, 2016]{chen2016xgboost}
Chen, T. and Guestrin, C. (2016).
\newblock Xgboost: A scalable tree boosting system.
\newblock In {\em Proceedings of the 22nd ACM SIGKDD International Conference
  on Knowledge Discovery and Data Mining}, KDD '16, page 785–794, New York,
  NY, USA. Association for Computing Machinery.

\bibitem[Crowdstrike, 2022]{crowdstrike}
Crowdstrike (2022).
\newblock Lateral movement explained: What is lateral movement?
\newblock
  \url{https://www.crowdstrike.com/cybersecurity-101/lateral-movement/}.
\newblock Accessed: 2022-08-13.

\bibitem[Du et~al., 2018]{du2018towards}
Du, M., Liu, N., Song, Q., and Hu, X. (2018).
\newblock Towards explanation of dnn-based prediction with guided feature
  inversion.
\newblock In {\em Proceedings of the 24th ACM SIGKDD International Conference
  on Knowledge Discovery and Data Mining}, KDD '18, page 1358–1367, New York,
  NY, USA. Association for Computing Machinery.

\bibitem[Friedman et~al., 2000]{10.1214/aos/1016218223}
Friedman, J., Hastie, T., and Tibshirani, R. (2000).
\newblock {Additive logistic regression: a statistical view of boosting (With
  discussion and a rejoinder by the authors)}.
\newblock {\em The Annals of Statistics}, 28(2):337--407.

\bibitem[Hartigan and Wong, 1979]{hartigan1979algorithm}
Hartigan, J.~A. and Wong, M.~A. (1979).
\newblock Algorithm as 136: A k-means clustering algorithm.
\newblock {\em Journal of the Royal Statistical Society. Series C (Applied
  Statistics)}, 28(1):100--108.

\bibitem[Hecht-Nielsen, 1992]{hecht1992theory}
Hecht-Nielsen, R. (1992).
\newblock Theory of the backpropagation neural network.
\newblock In {\em Neural Networks for Perception}, pages 65--93. Academic
  Press.

\bibitem[Hosmer~Jr et~al., 2013]{hosmer2013applied}
Hosmer~Jr, D.~W., Lemeshow, S., and Sturdivant, R.~X. (2013).
\newblock {\em Applied Logistic Regression}.
\newblock Wiley Series in Probability and Statistics. Wiley.

\bibitem[Ke et~al., 2017]{ke2017lightgbm}
Ke, G., Meng, Q., Finley, T., Wang, T., Chen, W., Ma, W., Ye, Q., and Liu,
  T.-Y. (2017).
\newblock Lightgbm: A highly efficient gradient boosting decision tree.
\newblock In {\em Advances in Neural Information Processing Systems},
  volume~30, pages 3146--3154.

\bibitem[Kent, 2015]{kent2015comprehensive}
Kent, A.~D. (2015).
\newblock {Comprehensive, Multi-Source Cyber-Security Events}.
\newblock Los Alamos National Laboratory.

\bibitem[Kent, 2016]{lanl-data}
Kent, A.~D. (2016).
\newblock {\em Cyber security data sources for dynamic network research},
  volume~1, chapter~2, pages 37--65.
\newblock World Scientific.

\bibitem[Lundberg and Lee, 2017]{shap}
Lundberg, S.~M. and Lee, S. (2017).
\newblock A unified approach to interpreting model predictions.
\newblock In {\em Advances in Neural Information Processing Systems}, pages
  4765--4774.

\bibitem[McInnes et~al., 2018]{umap}
McInnes, L., Healy, J., and Melville, J. (2018).
\newblock Umap: Uniform manifold approximation and projection for dimension
  reduction.

\bibitem[Mossburg et~al., 2016]{deloitte-cyberattack}
Mossburg, E., Gelinne, J., and Calzada, H. (2016).
\newblock Beneath the surface of a cyberattack.
\newblock
  \url{https://www2.deloitte.com/content/dam/Deloitte/us/Documents/risk/us-risk-beneath-the-surface-of-a-cyber-attack.pdf}.
\newblock Accessed: 2022-08-13.

\bibitem[Rocklin, 2015]{rocklin2015dask}
Rocklin, M. (2015).
\newblock Dask: Parallel computation with blocked algorithms and task
  scheduling.
\newblock In {\em Proceedings of the 14th Python in Science Conference}, pages
  126 -- 132.

\bibitem[Short and Fukunaga, 1981]{1056403}
Short, R. and Fukunaga, K. (1981).
\newblock The optimal distance measure for nearest neighbor classification.
\newblock {\em IEEE Transactions on Information Theory}, 27(5):622--627.

\bibitem[Song and Lu, 2015]{song2015decision}
Song, Y. and Lu, Y. (2015).
\newblock Decision tree methods: applications for classification and
  prediction.
\newblock {\em Shanghai Archives of Psychiatry}, 27(2):130--135.

\bibitem[Uppstr{\"o}mer and R{\aa}berg, 2019]{uppstromer2019detecting}
Uppstr{\"o}mer, V. and R{\aa}berg, H. (2019).
\newblock Detecting lateral movement in microsoft active directory log files: A
  supervised machine learning approach.
\newblock Master's thesis, Faculty of Computing, Blekinge Institute of
  Technology, SE-371 79 Karlskrona, Sweden.

\bibitem[Wold et~al., 1987]{wold1987principal}
Wold, S., Esbensen, K., and Geladi, P. (1987).
\newblock Principal component analysis.
\newblock {\em Chemometrics and Intelligent Laboratory Systems}, 2(1):37--52.

\bibitem[Zhang and Ma, 2012]{zhang2012ensemble}
Zhang, C. and Ma, Y. (2012).
\newblock {\em Ensemble Machine Learning: Methods and Applications}.
\newblock Springer, New York, NY, USA.

\end{thebibliography}
\end{document}